Original Manuscript

Title:

**Cognitive computation of brain disorders based primarily on ocular responses**

Authors:

Xiaotao Li, [*1,2] Xuejing Chen,[3] Fangfang Fan,[4] Li Ning[5], Kangguang Lin[6,7], Zan Chen[1], Zhenyun Qin[8], Albert S. Yeung[9], Liping Wang[1], Xiaojian Li [1], Kwok-Fai So,[7,10]

1 Brain Cognition and Brain Disease Institute, Shenzhen Institutes of Advanced Technology, Chinese Academy of Sciences; Shenzhen-Hong Kong Institute of Brain Science-Shenzhen Fundamental Research Institutions, Shenzhen 518055, Guangdong Province, China. 2 Department of Brain and Cognitive Sciences, Massachusetts Institute of Technology, Cambridge MA, USA. 3 Retina Division, Department of Ophthalmology, Boston University Eye Associates, Boston University, Boston MA, USA. 4 Department of Neurology, Harvard Medical School, Harvard University, Boston MA, USA. 5 Center for High Performance Computing, Shenzhen Institutes of Advanced Technology, Chinese Academy of Sciences, Shenzhen, China. 6 Department of Affective Disorders and Academician Workstation of Mood and Brain Sciences, The Affiliated Brain Hospital of Guangzhou Medical University (Guangzhou Huiai Hospital), Guangzhou, China. 7 Guangdong-Hong Kong-Macau Institute of CNS Regeneration, Jinan University, Guangzhou, China. 8 School of Mathematical science and Key Laboratory for Nonlinear Mathematical Models and Methods, Fudan University, Shanghai, China. 9 Depression Clinical and Research Program, Department of Psychiatry, Massachusetts General Hospital, Boston MA, USA. 10 Department of Ophthalmology and The State Key Laboratory of Brain and Cognitive Sciences, University of Hong Kong, Hong Kong.

*Corresponding author. Email address: xtli@mit.edu

**Abstract**

The present review presents multiple techniques in which ocular assessments may serve as a noninvasive approach for the early diagnoses of various cognitive and psychiatric disorders, such as Alzheimer's disease (AD), autism spectrum disorder (ASD), schizophrenia (SZ), and major depressive disorder (MDD). Real-time ocular responses are tightly associated with emotional and cognitive processing within the central nervous system. Patterns seen in saccades, pupillary responses, and blinking, as well as retinal microvasculature and morphology visualized *via* office-based ophthalmic imaging, are potential biomarkers for the screening and evaluation of cognitive and psychiatric disorders. Additionally, rapid advances in artificial intelligence (AI) present a growing opportunity to use machine-learning-based AI, especially deep-learning neural networks, to shed new light on the field of cognitive neuroscience, which may lead to novel evaluations and interventions *via* ocular approaches for cognitive and psychiatric disorders.

## Introduction

The neurosensory retina places a critical role to the functioning of our central nervous system (CNS), the latter of which processes our sensory input, motor output, emotions, cognition, and even consciousness[1]. Multiple studies have shown that ocular evaluations can be used to assess neurologic and psychiatric disorders[2]. Many neurological and psychiatric disorders — such as glaucoma, stroke, Parkinson's disease (PD), autism spectrum disorder (ASD), Alzheimer's disease (AD), major depressive disorder (MDD), and schizophrenia (SZ) — lead to considerable personal suffering, financial costs, and social burden[3]; fortunately, distinct ocular findings have exhibited the possibility of ocular assessments as biomarkers for these disorders[2]. Since neurological disorders represent one of the most challenging issues in modern society, novel approaches are needed to advance psychiatric medicine, especially in terms of objective cognitive measurements[4] and real-time interventions for cognitive deficits[5].

Recently, the DeepMind or Google was able to detect retinal diseases and cardiovascular risk factors using artificial intelligence (AI) algorithms on retinal images[6,7]. Early AI studies have already shown that AI-based disease detection is possible, and the potential of AI to impact the next generation of medical care as well[8]. Given the increasing data connecting ocular parameters with cognitive and psychiatric disease states, it is possible that applying AI algorithms on ocular parameters would be helpful for the detection and evaluation of these diseases, especially with the rapid advancement of computer vision (CV) and deep-learning algorithms[9]. In the present review, we discuss the novel approach of using AI to evaluate human affective and cognitive states based on real-time ocular responses.

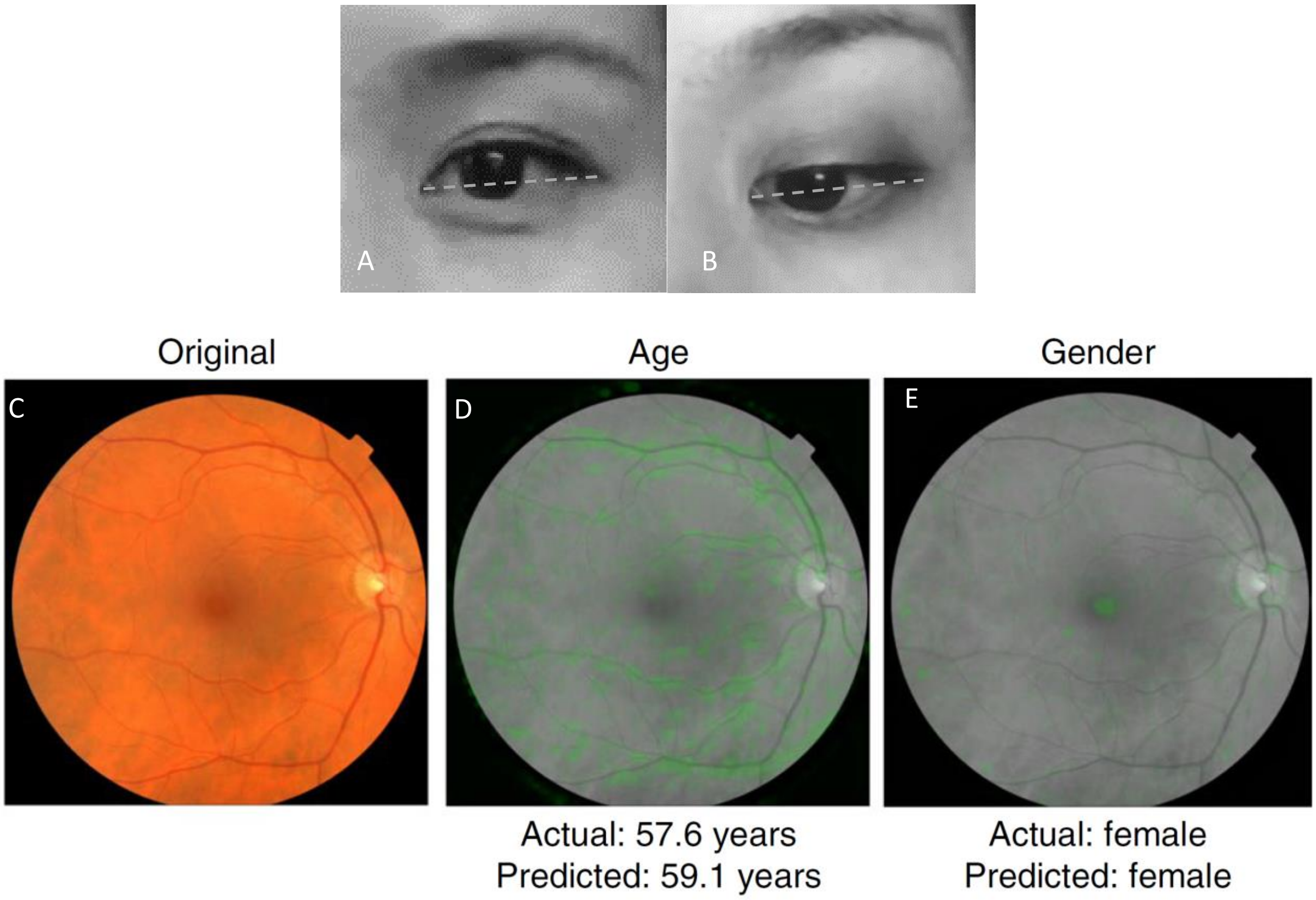

**Figure 1 The eye as a window to uncover healthy level.** (A) A restful and calm eye (positive state) is shown compared to (B) a stressful and anxious eye (negative state). Note that a positive state is more frequently associated with an upside view of eyes, whereas a negative state exhibits a more downside view of eyes. The eye images shown here are presented following permission from the corresponding subjects. (C) Example of a retinal fundus image in color, whereas (D) and (E) show the same retinal image, but in black and white. Predictions of age and gender mainly rely on the features of the vasculature and optic disc, as indicated by the green color in these images. The images in (C)–(E) were adapted from Poplin et al[7] (will get the permission from them).

**Eye-brain connections**

The eyes and the brain are intimately connected. Approximately 80% of the sensory input to the human brain originates from the visual system, which begins at the retina[10]. The axons of retinal ganglion cells (RGCs) send visual stimuli collected on the neurosensory retina to the CNS[2]. In the mammalian brain, there are approximately 20 nuclei that receive projections from the retina; for example, the lateral geniculate nucleus serves as a thalamic visual relay to primary visual cortex, the superior colliculus is responsible for visuomotor processing, and the hypothalamic suprachiasmatic nucleus is involved in non-visual hormonal photoentrainment[11].

The eyes are also often linked to facial expressions. For example, eye-widening can be a sign of fear, whereas eye narrowing can be a sign of disgust[10]. Interestingly, Lee et al. demonstrated that eye-widening enhances one's visual field, thereby improving stimulus detection, while eye narrowing increases visual acuity, which improves objective discrimination[12,13]. Moreover, visual attention, pupillary responses, and spontaneous blinking are regarded as non-invasive and complementary measures of cognition[14]. Vision-based attentional control includes the planning and timing of precise eye movements, which has been shown to be controlled by neural networks spanning cortical, subcortical, and cerebellar areas that have been extensively investigated in both humans and non-human primates[15–17] (Figure 2). Pupillary responses are primarily modulated by norepinephrine in the locus coeruleus, which controls physiological arousal and attention, and can be used as a measure of subjective task difficulty and mental effort. In contrast, spontaneous eye-blink rates are tightly correlated with CNS dopaminergic levels and are associated with processes underlying learning and goal-directed behavior[14]. Below, we outline in detail the connections between ocular assessments and some psychiatric disorders including Alzheimer's disease (AD), autism spectrum disorder (ASD), schizophrenia (SZ) and major depressive disorder (MDD).

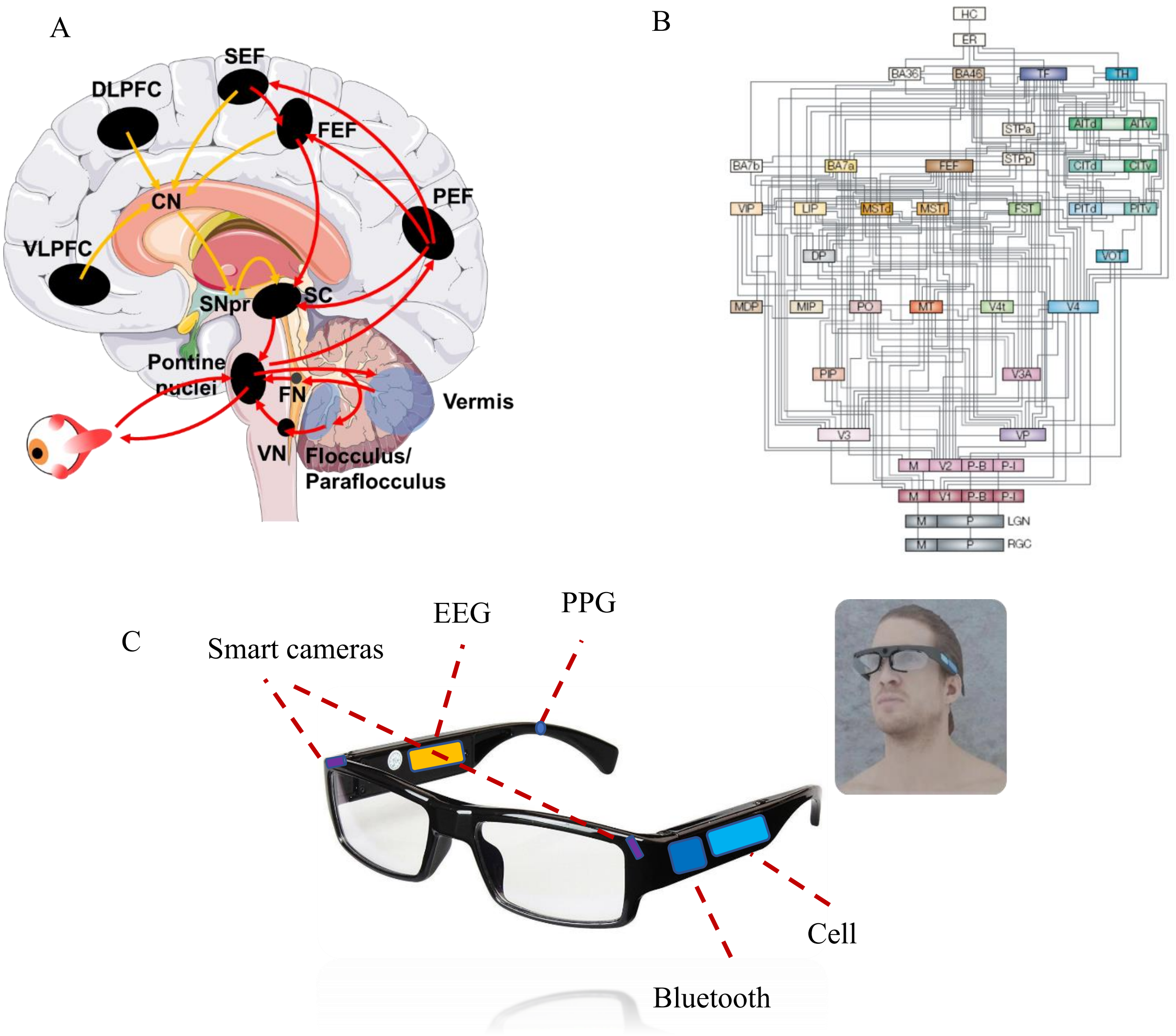

**Figure 2 Brain circuits involved in visually guided cognitive function and their use in novel applications.** (A) Complex neural networks spanning cortical, subcortical, and cerebellar areas are involved in voluntary saccadic eye movements for attentional control. The image was modified from Johnson et al.[18]. Red arrows indicate the direct pathway (PEF, the parietal eye fields; FEF, frontal eye field; SEF, supplementary eye field) to the SC and brainstem premotor regions, while yellow arrows indicate the indirect pathway to the SC and brainstem premotor regions via the basal

ganglia (striatum, subthalamic nucleus, globus pallidus, and substantia nigra pars reticularis). (B) An architectural model of the hierarchy of visual cortical circuitry, adapted from Felleman and Van[19]. There is a feedforward ascending pathway of the vision system from the retina to cortex, as well as a feedback descending pathway from the cortex to multiple downstream areas. (C) A potential application of eye-brain engineering developed to compute human brain disorders mainly based on ocular responses, combined with other biological signals including electroencephalography (EEG) and photoplethysmography (PPG).

**Alzheimer's disease (AD)**

Ocular assessments of AD patients have demonstrated saccadic dysfunctions indicative of poor visual attention. In particular, AD patients have difficulty focusing on fixed objects[20]. Prettyman et al. showed in 1997 that there was a 75% greater latency in pupillary constriction in AD patients compared with that in age-matched controls.[22] Additionally, compared with that in age-matched controls, AD patients have markedly decreased visual contrast sensitivity, which is evident even at early stages of AD.[24] Patients with AD also have altered retinal microvasculature, such as sparser and more tortuous retinal vessels and narrower retinal venules,[25] and decreased retinal blood flow and blood-column diameter, as indicated by laser Doppler flowmetry[21]. Studies using optical coherence tomography (OCT) have shown a gradual decrease in retinal nerve fiber layer thickness (RNFL), most prominently in the superior quadrants, when comparing patients with no AD to mild AD to severe AD.[22,23]

In summary, AD patients exhibit a number of specific ocular findings ranging from ocular movements, pupillary responses, contrast sensitivity, and retinal microvasculature and morphology. These overlapping ocular findings provide an opportunity for their collective use as biomarkers for machine-learning algorithms for indicating AD.[29]

**Autism spectrum disorder (ASD)**

A lack of normal eye contact during social interactions is one of the main clinical features of autism spectrum disorder (ASD).[24] Screening for ocular fixation at 2–6 months old can provide early detection and even interventions for children with ASD.[31] Full-field electroretinograms (ERGs), which measure specific cellular functions within the retina, exhibit decreases in rod b-wave amplitude in ASD individuals[25].

Additionally, different types of oculomotor dysfunction—such as saccade dysmetria (over- or under-shooting of visual targets), loss of saccadic inhibition, and fixation impairment—have all been shown in patients with ASD[18]. Saccade dysmetria may be secondary to dysfunction of neural networks connecting cerebellar vermal-fastigial circuitry to brainstem premotor nuclei that regulate oculomotor movements[26,27] (Figure 2A). A diminished ability to inhibit reflexive saccades during anti-saccade tasks is thought to be associated with the characteristic repetitive behaviors seen in patients with ASD and may also be associated with hyperactivation of the frontal eye field (FEF) and anterior cingulate cortex (ACC), both of which are involved in conflict monitoring.[28,29] Lastly, fixation is often significantly impaired in ASD patients, which is possibly secondary to decreased activation of V5 and a reduced top-down modulation of sensorimotor processing[30].

**Schizophrenia (SZ)**

Abnormal retinal findings in patients with SZ include dilated retinal venules, RNFL thinning, and ERG abnormalities[31,32]. A twin study showed a positive correlation between wider retinal venules and more severe psychotic symptoms[33], suggesting the possible use of retinal venule diameter as a biomarker for SZ. RNFL thinning, which corresponds to the loss of RGCs axons, is seen in patients with SZ[34,35] and also in patients with PD[36] and AD[37]. ERG abnormalities in SZ indicate reduced functionalities of rod and cone photoreceptors, bipolar cells, and RGCs, all of which can reflect the deregulation of neurotransmitters, such as dopamine, in the neurosensory retina that is a part of the CNS[38,39]. Furthermore, portal ERG devices exist and may be used in psychiatry clinics for screening and evaluation of SZ [40].

**Major depressive disorder (MDD)**

Reduced contrast sensitivity is often seen in individuals with MDD, both medicated and unmedicated[41]. There is even a strong correlation between contrast gain and depression severity, as indicated by recorded ERG patterns[41,42]. When compared with those of healthy controls, patients with MDD often have higher error rates and increased reaction times in performing anti-saccade tasks[43,44]. Additionally, patients with melancholic depression, when compared with parameters in healthy controls and non-melancholic depressed patients, exhibit primary saccades with longer latencies, reduced peak velocities, and greater hypometricity during saccadic eye movement tasks[45].

Patients with seasonal affective disorder (SAD) have a significantly reduced post-illumination pupillary response (PIPR) and also a lower PIPR percent change in response to blue-light stimuli, as demonstrated by infrared pupillometry[46,47]. These characteristics are most likely to be associated with dysfunction of melanopsin-expressing intrinsically photosensitive RGCs (ipRGCs), since ipRGCs often contribute to pupillary response function, particularly during sustained-state pupillary constriction[48,49]. However, the PIPR in response to blue-light stimulation changed only in SAD patients carrying the OPN4 I394T genotype[46].

Table 1 **Multiple changes in ocular parameters *via* ophthalmological assessments are associated with neurological disorders.** Some similar features among these diseases further indicate a requirement of more precise analyses *via* machine learning and deep learning.

| | Saccades | Pupillary reflexes | RNFL | Microvasculature | ERG |
|---|---|---|---|---|---|
| Alzheimer's disease | Poor eye fixation[20]<br><br>Decrease eye fixation | Delayed pupillary constriction[50,51] | Reduced RNFL thickness especially in the superior quadrant[24,25] | Narrower retinal venules and sparser and more tortuous retinal vessels[52] | Markedly decreased contrast sensitivity[53]<br><br>Decreased rod b- |
| Autism | at 2–6 months old[54]; saccade dysmetria[18]; impaired tracking of moving targets[30] | --- | --- | --- | wave amplitude in flash ERG[25] |
| Schizophrenia | --- | --- | Thinning of RNFL[34,35] | Widened retinal venules[33] | Abnormal ERG amplitudes including rods, cones, bipolar cells, and RGCs[42,43,40] |
| Major depression | Elevated error rates and increased reaction times[43,44] | Reduced PIPR and a lower PIPR percent change in response to blue light in patients with SAD[46] | --- | --- | Significantly reduced contrast sensitivity using pattern ERG[41,42] |

**Computer vision (CV)**

As previously outlined, cognitive and psychiatric disorders often have ocular manifestations that may be shared amongst multiple diseases or may be specific to a singular disease (Table 1). Vast psychological and economic burdens caused by neurological disorders call for more precise analyses of these disorders via machine learning and deep-learning algorithms.

CV is regarded as one of the most powerful tools to push AI applications into healthcare areas, as it exhibits a high capability of auto-screening diseases such as skin cancer[55] and diabetic retinopathy[8,56]. Following a rapid development of deep-learning-based AI, CV now exhibits an impressive resolution that is close to that of human vision and maybe beyond sometime. Hence, CV is likely to drive AI to provide diagnostic tools for neurological disorders, as machines have now been able to be trained to read human emotional and cognitive states, especially in terms of automated detection of facial expressions such as fear and fatigue. Individuals with psychiatric disorders may benefit from CV-based AI applications in neurological healthcare, particularly to aid in patient self-monitoring of symptoms and in conducting real-time interventions for recovery of social and psychological abilities. Furthermore, CV-based AI recognition of human emotional and cognitive states may be more precisely achieved with automated detection and analyses of ocular responses. Nevertheless, currently available CV used for facial-expression recognition often neglects human eye movement and other primary ocular parameters, which usually contain abundant information of human affective states. Since Google AI applications focused on retinal and eye features have already demonstrated that this approach noninvasively and conveniently yields accessible and useful health information[6,8], such eye-focused CV-based AI should be explored further for clinical applications in future. Hence, the eye as a window into the brain will be able to be used to obtain useful health information, not only for eye diseases but also for determining cardiovascular risk factors[7] and brain cognitive states (Table 1).

## Discussion

Medical issues involving challenging human diseases are often tightly associated with big data, and AI algorithms have been demonstrated to leverage such big data to aid in solving these medical issues[9]. Although AI applications in medicine are only at early stages, AI-based automated diagnoses have already contributed to the identification of several types of cancers and retinal diseases[55,8,56]. Many reports on AI practices engaged in medical diagnoses have involved image recognition via supervised learning using deep neural networks, which has helped in effectively interpreting cancer slides[57], retinal images[58], and brain scans[59]; however, many of these applications have only been completed at preliminary stages and have not undergone formal peer-reviewed publication. AI performance has been frequently leveraged in ophthalmology since retinal images are relatively easy to obtain using fundus imaging or OCT without any invasion. Additionally, diagnostic standards of eye diseases have become more well-defined. Many eye diseases—including diabetic retinopathy[60], age-related macular degeneration[61], and congenital cataracts[62]—have already been assessed via deep-learning neural networks, and the majority of these applications have exhibited remarkable accuracies comparable to those of eye specialists[9]. However, brain disorders tend to be more complicated than other medical issues and often lead to a greater burden to human society, as demonstrated by at least 350 million people currently suffering from major depression in the world[63]. There is a huge potential for AI to lend its power to patients with affective disorders and the caregivers helping these patients. Specifically, the neurosensory retina, a key component of the eye and vision, is a direct embryological extension of the CNS[1]. Therefore, the utility of AI algorithms to detect abnormal ocular responses associated with brain disorders is worth testing for its feasibility and efficacy. CV-based AI algorithms for automated detection and analyses of ocular responses may represent promising tools for noninvasively detecting differences in ocular responses[10], particularly those associated with brain disorders such as AD, ASD, SZ and MDD. Nevertheless, currently available datasets on emotional recognition—such as JAFFE, FERA, and $CK^+$ — have usually been based on thousands of facial images captured in the laboratory but often lack more information about human emotions and cognitive states in the natural environment. In order to achieve reliable detection of human emotional and cognitive states in real time for brain disorders, a more quantitative capability of emotional recognition *via* AI algorithms is necessary, which will require more dynamic features including eye movements captured from the natural environment.

Undoubtedly, AI has begun to shed new light on brain disorders. Cognoa applied clinical data from thousands of children at risk for ASD in order to train and develop an AI platform, which may provide earlier diagnostics and personalized therapeutics for autistic children[64] and was approved by the FDA in 2018. Also, Rudovic et al developed a personalized machine-learning framework with deep-learning algorithms for further automated detection of affective characteristics and engagement in children during robot-assisted therapy for autism[65]. In terms of auto-screening depression, Alhanai et al used audio and text features to train a neural network with long short-term memory, which was found to be comparable to traditional evaluations *via* depression questionnaires[66]. Haque et al trained their AI using spoken language and 3D facial expressions commonly available in smartphones to measure depression severity, which yielded an 83.3% sensitivity and 82.6% specificity[67]. However, the resolution of 3D facial scans was too low to resolve ocular parameters[67]. A project in China that our team is currently running is aimed at developing an AI platform that utilizes ocular data to train a model to detect brain states under natural conditions. This AI platform is mainly based on real-time ocular responses and may be able to determine brain emotional and cognitive states of individuals with brain disorders. This core function will be obtained accessibly through either an ordinary smartphone or wearable smart glasses, the latter of which we are currently designing.

Some details regarding AI implementation in healthcare still require consideration. The first crucial issue is how to effectively collect big data with a high quality of valid features for AI algorithms. In terms of AI recognition of facial expression, high-resolution imaging is often required, especially for the application of neurological disorders. Multiple ocular data should not be neglected for emotional recognition since eye expression plays a key role in social communication. Parameters such as eye movement, pupillary responses, and blinking rates (Table

1) contain detailed and fruitful information on human affective states[14]. Micro-expression detection is also sometimes required for ground truth of facial expression[68]. Some key features are unable to be shown by one single image and instead require dynamic videos with high framerates. In addition, high-resolution imaging is potentially beneficial for obtaining more healthcare data via PPG[69,70], such as heart rate variability[72]. Another key issue is in terms of privacy protection when facial data are obtained to develop AI algorithms[71]. First, participants need to be provided informed consent regarding data collection, and investigators need to advise participants of their rights and summarize what is expected for participation in the corresponding study[71]. In our pilot design, all facial or ocular data will be obtained and stored by participants themselves through their mobile phones, and they have the right to decide if data are shared without identifying their personal information. In addition, some portable or wearable devices have been developed for brain healthcare since they are better for early detection of symptoms involved in brain disorders, rather than diagnosing symptoms at the later and severe stages, and other technologies exhibit this utility, such as functional MRI for evaluating depression[72], or computed tomography for head-trauma screening[73]. The wearable smart glasses that we designed (Figure 2C, the related patent is in the process) are able to collect many different parameters involving ocular responses, such as saccades, pupillary responses, and blinking parameters. Additionally, other biological data may be considered as well, such as EEG and PPG (Figure 2C). We have defined an integrated eye-brain engineering tool for human state recognition that may enable powerful detection and evaluation of brain states in real time via machine learning.

As AI is still at the early stage of being integrated into neurological healthcare, integration of human biological intelligence (BI) and machine-learning-based AI will also need to be conducted. AI alone is known to be insufficient for detecting neurological disorders since machine learning is entirely dependent on the availability of collected data. The quality of collected data is vital, that will require the use of BI including related knowledge, information and statistics to be involved. Current AI detection of facial expressions is often only achieved at a qualitative level and has been categorized into seven basic expressions with some composite expressions[74], with an accuracy ratio of less than 70%. Therefore, AI detection of facial expressions requires improvement to a quantitative level via more BI, especially in the context of cognitive neuroscience. Obtaining fruitful information on ocular responses with advanced devices and AI technology, we can perhaps learn more clearly about the threshold values of brain disorders distinguished from normal brain function and learn more about specific features of different brain disorders. These efforts may be beneficial for conducting effective and timely prevention and treatment for brain disorders. To this aim, machine-learning-based AI, especially deep-learning neural networks, will likely be instrumental in further advancing clinical applications in neuroscience[75].

In general, brain disorders can be assessed by ocular detection while that certainly needs to consider the exclusion of the eye disease situation and that well-trained AI will offer its support again[4]. Eventually, advanced AI will help those patients with brain disorders directly and currently, differing from gene therapy that plans to offer new prospective for the next generation of psychiatric patients but not for current ones. It can be achieved mainly through auto-detection with AI algorithms, self-evaluation with wearable sensors, and timely-intervention with human-computer interface (HCI) as well. May the force of AI be with patients of brain disorders particularly using a promising approach to detect ocular responses in the real time.

## Acknowledgments

This article was supported by grants from the Commission on Innovation and Technology in Shenzhen Municipality of China (JCYJ20150630114942262), the International Postdoctoral Exchange Fellowship Program 2016 by the Office of China Postdoctoral Council (20160021), the Hugo Shong fellowship supporting postdoctoral research at the McGovern Institute for Brain Research at MIT, the National Key R&D Program of China (2017YFC1310503), Key-Area Research and Development Program of Guangdong Province (2018B030331001), and the Hong Kong, Macao & Taiwan Science and Technology Cooperation Innovation Platform in Universities in Guangdong Province (2013gjhz0002). In particular, we would like to thank the Venture Mentoring Service (VMS) at MIT, Cambridge, MA, USA.

Ref: